\documentclass[a4paper,showpacs,preprintnumbers,amsmath,amssymb,twocolumn,pre,nofootinbib]{revtex4}
\usepackage{graphicx,lscape,float,dcolumn,bm,url}
\graphicspath{{./figs/}}

\begin{document}

\newcommand{\Rey}{\mathrm{Re}}
\newcommand{\ReL}{\mathrm{Re}}
\newcommand{\ff}{\bm{f}}
\newcommand{\kk}{\bm{k}}
\newcommand{\eee}{\bm{e}}
\renewcommand{\vec}{\bm}
\newcommand{\pp}{^{\scriptscriptstyle\mathord{+} }}

\title{The ``zeroth law'' of turbulence:
        Isotropic turbulence simulations revisited}
\author{Bruce R.\ Pearson}
  \email{Bruce.Pearson@nottingham.ac.uk}
  \affiliation{School of Mechanical, Materials,
        Manufacturing Engineering \& Management.\\
        University of Nottingham, Nottingham NG7 2RD, UK.}
\author{Tarek A.\ Yousef}
  \email{Tarek.Yousef@mtf.ntnu.no}
  \affiliation{
        Dept.~of Energy and Process Engineering,
        The Norwegian University of Science and Technology,
        Kolbj{\o }rn Hejes vei 2B, N-7491 Trondheim, Norway}
\author{Nils Erland L.\ Haugen}
  \email{Nils.Haugen@phys.ntnu.no}
  \affiliation{Dept.~of Physics, The Norwegian University of Science
        and Technology,\\ H{\o}yskoleringen 5, N-7034 Trondheim, Norway}
\author{Axel Brandenburg}
  \email{Brandenb@nordita.dk}
  \affiliation{NORDITA, Blegdamsvej 17, DK-2100 Copenhagen \O, Denmark}
\author{Per-\AA ge Krogstad}
  \email{Per.A.Krogstad@mtf.ntnu.no}
  \affiliation{
        Dept. of Energy and Process Engineering,
        The Norwegian University of Science and Technology,
        Kolbj{\o }rn Hejes vei 2B, N-7491 Trondheim, Norway}

\date{\today,  CVS ${}$Revision: 1.69 ${}$}

\begin{abstract}
The dimensionless kinetic energy dissipation rate $C_{\varepsilon }$ is estimated from numerical simulations of
statistically stationary isotropic box turbulence that is slightly compressible. The Taylor microscale Reynolds number
$(\Rey_{\lambda})$ range is $20\lesssim \Rey_{\lambda }\lesssim 220$ and the statistical stationarity is achieved with
a random phase forcing method. The strong $\Rey_{\lambda }$ dependence of $C_\varepsilon$ abates when $\Rey_{\lambda }
\approx 100$ after which $C_\varepsilon$ slowly approaches $\approx 0.5,$ a value slightly different to previously
reported simulations but in good agreement with experimental results. If $C_\varepsilon$ is estimated at a specific
time step from the time series of the quantities involved it is necessary to account for the time lag between energy
injection and energy dissipation. Also, the resulting value can differ from the ensemble averaged value by up to $\pm
30\%.$ This may explain the spread in results from previously published estimates of $C_\varepsilon.$
\end{abstract}

 \maketitle

\section{Introduction}
The notion that the mean turbulent kinetic energy dissipation rate $\varepsilon$ is finite and independent of
viscosity $\nu$ was originally proposed by G.~I. Taylor\cite{t35}. Its importance is so recognized now that it is
commonly referred to as the ``zeroth law'' of turbulence. Its existence was assumed by von K\'{a}rm\'{a}n and
Howarth, Loitsianskii and also, significantly, Kolmogorov\cite{k41} in establishing his celebrated similarity
hypotheses for the structure of the inertial range of turbulence. Kolmogorov assumed the small scale structure of
turbulence to be locally isotropic in space and locally stationary in time - which implies the equality of
turbulent kinetic energy injection at the large scales with the rate of turbulent kinetic energy dissipation at the
small scales\cite{ccddt97}. Although this view should be strictly applied only to steady turbulence, the mechanism
of the dissipation of turbulent kinetic energy can be considered the most fundamental aspect of turbulence not only
from a theoretical viewpoint but also from a turbulence modeling viewpoint. Indeed, the mechanism that sets the
level of turbulent dissipation in flows that are unsteady is a difficult, if not intractable, aspect of turbulence
modeling.

The rate of turbulent kinetic energy dissipation is determined by the rate of energy passed from the large-scale
eddies to the next smaller scale eddies via a forward cascade until the energy is eventually dissipated by
viscosity. Thus, $C_\varepsilon$ defined as,
\begin{equation}
C_{\varepsilon}= \varepsilon  L/{u'}^3 , \label{eq:ceps}
\end{equation}
(here, $ \varepsilon $ is the mean energy dissipation rate per unit mass, $L$ and $u'$ are characteristic large length
and velocity scales respectively) should be independent of the Reynolds number and of order unity. An increase in
Reynolds number should only result in an increase in the typical wave number where dissipation takes place\cite{l92}.
In the past few years there have been a number of numerical (see Ref.~\cite{kiyiu03} and references therein) and
experimental (see Refs.~\cite{pkw02,w02,pkj03} for recent results) efforts to determine the value of $C_\varepsilon$
and its dependence on the Reynolds number. Perhaps the most convincing of these are the numerical attempts since there
is no re-course to one-dimensional surrogacy as there is for experiments. Notwithstanding this fact, there is good
agreement, both numerically and experimentally, with the long held view that $C_\varepsilon$ is $\sim O(1)$ when the
Reynolds number is sufficiently high. The collection of isotropic simulation results for $C_\varepsilon$ shown in Ref.
\cite{kiyiu03} indicates that ``high enough'' Reynolds number appears to be $\Rey_\lambda \sim O(100).$ Here,
$\Rey_\lambda (={u'}^2[15/\nu\varepsilon ]^{1/2})$ is the Taylor microscale Reynolds number. At higher $\Rey_\lambda$,
slow $\Rey_\lambda$ dependencies for $C_{\varepsilon}$, such as that proposed by Lohse\cite{l94}, cannot be ruled out.
Measuring such $\Rey_\lambda$ dependencies, either numerically or experimentally, will be close to impossible.

One unresolved issue is that raised by Sreenivasan\cite{s98}. After assembling all the then known experimental decaying
grid turbulence data\cite{s84} and numerical data for both decaying and stationary isotropic turbulence he concludes
that ``the asymptotic value (of $C_\varepsilon$) might depend on the nature of large-scale forcing, or, perhaps, on the
structure of the large scale.'' He also demonstrates\cite{s95b} in homogeneously sheared flows that the large structure
does influence $C_\varepsilon.$ However, it might be argued that these results were obtained at low Reynolds numbers
and the issue of a universal asymptotic value for $C_\varepsilon$ could still be considered open. Alternatively it
could be argued that homogeneous shear flows and the like are strictly unsteady turbulent flows and the zeroth law, in
its simplest guise, should not be expected to apply to such flows. The possibility of some characteristics of
large-scale turbulence being universal should not be ruled out. The recent observation that input power fluctuations,
when properly re-scaled, appear universal\cite{bhp98} may be construed to suggest the possibility of universality for
$C_{\varepsilon}$. The aim of the present work is to estimate $C_\varepsilon$ from direct numerical simulations (DNS)
of statistically stationary isotropic turbulence and compare with previously reported DNS results (summarized in Fig.~3
of Ref.~\cite{kiyiu03}) and experiments carried out in regions of low $\left (dU/dy \approx dU/dy|_{\rm max}/2 \right
)$ or zero mean shear. The present DNS scheme differs from methods already reported in that a high-order finite
difference method is used. To our knowledge, these are the first finite difference results for $C_\varepsilon.$ Hence,
it is worthwhile to test if different numerics and forcing at the large scales result in vastly different values for
$C_\varepsilon$ to those already reported.

\begin{table*}
\begin{ruledtabular}
\begin{tabular}{lccccccccccccc}
  Run & $N$ & $\Rey_\lambda$ & $T_{\rm tot}/T$ &
  $\nu \left(\times 10^{4}\right )$ & $ \varepsilon
  \left(\times 10^{5}\right )$&$\Delta t/t_\kappa\left(\times 10^{2}\right )$  &$L$ & $\lambda$ & $u'\left(\times 10^{2}\right )$ & $\tau_{\rm max}/T$ & $C_\varepsilon$ &
  $\eta \left(\times 10^{3} \right )$ & $k_{\max} \eta$ \\ \hline
  A & 32 &  20 &  31 &  40  & 24 & 1.9  &1.9 &  1.2 &  7.1 &  0.15 &  1.2 & 128 & 2.1\\ 
  B & 64 &  42 &  30 &  15  & 22 & 1.5  &1.6 & 0.81 &  7.8 &  0.37 & 0.75 &  63 & 2.0\\ 
  C &128 &  90 &  11 &  4.0 & 24 & 1.5  &1.3 & 0.43 &  8.4 &  0.62 & 0.54 &  23 & 1.5\\ 
  D &256 &  92 &  19 &  4.0 & 21 & 0.71 &1.4 & 0.45 &  8.1 &  0.69 & 0.53 &  24 & 3.0\\ 
  E &256 & 152 &  20 &  1.6 & 21 & 1.1  &1.4 & 0.29 &  8.4 &  0.74 & 0.49 &  12 & 1.5\\ 
  F &512 & 219 &   7 &  0.80& 25 & 0.86 &1.3 & 0.20 &  8.9 &  0.67 & 0.47 &   7 & 1.7\\ 
\end{tabular}
\end{ruledtabular}
  \caption{Examples of DNS parameters and average turbulence characteristics. $N$ is the number of grid points in each of
  the Cartesian directions, $\Rey_\lambda$ is the Taylor microscale Reynolds number $\equiv  u' \lambda/\nu$,
  $T_{\rm tot}$ is the total run time after the run became statistically stationary,
  $T$ is the eddy turnover time $\equiv L/u'$, $\Delta t$ is the run time increment, $t_\kappa$ is the Kolmogorov
  time scale $\equiv \nu ^{1/2} \varepsilon ^{-1/2},$
  $\lambda$ is the Taylor microscale $\equiv u'\sqrt{15 \nu/\varepsilon},$
  $\tau_{\rm max}$ is the average time for the energy cascade from large to small scales,
  and $\eta$ is the Kolmogorov length scale $\equiv \nu ^{3/4} \varepsilon ^{-1/4}$.}  \label{tab:average results}
\end{table*}

\section{Numerical Methods}
The data used for estimating $C_\varepsilon$ are obtained by solving the Navier Stokes equations for an isothermal
fluid with a constant kinematic viscosity $\nu$  and a constant sound speed $c_{\rm s}$. The governing equations are
given by
\begin{eqnarray}
\left(\partial_t+\vec u\cdot\nabla\right)\vec u=
-c_{\rm s}^2\nabla\ln\rho+ \vec f_{\mathrm{visc}} + \vec f \\
\left(\partial_t+\vec u\cdot\nabla\right) \ln\rho=
-\nabla\cdot\vec u.
\end{eqnarray}
The viscous force is
\begin{equation}
\vec f_{\mathrm{visc}}=\nu\left(\nabla^2\vec u+\textstyle{\frac13}\nabla\nabla\cdot\vec u
+2\nu\bm{\mathsf{S}}\cdot\nabla\ln\rho
 \right),
\end{equation}
where $\mathsf{S}_{ij}=\frac 12 (u_{i,j}+u_{j,i}) - \frac 13 \delta_{ij}\nabla\cdot\vec u$ is the traceless rate of strain
tensor. In the numerical simulations the system is forced (stirred) using random transversal waves given by
\begin{equation}
\vec f(\vec x,t) =f_0\vec e\cos\left[{\rm i}\vec k(t)\cdot\vec x +{\rm i}\phi(t)\right],
\label{nohel_forcing}
\end{equation}
where $\kk(t)$ is a wave number with magnitude between 1 and 2, while $\phi(t)$ is a phase between $-\pi$ and $\pi$.
Both $\phi(t)$ and $\vec k(t)$ are chosen randomly at each time step giving a forcing that is delta-correlated in time.
The random unit vector $\vec e$ is perpendicular to $\vec k$ and the forcing amplitude $f_0$ is  chosen such that the
root mean square Mach number for all runs is between $0.13$ and $0.15$ which is not too dissimilar to that found in the
wind-tunnel experiments to be discussed in the next section. For these weakly compressible simulations, the energies of
solenoidal and potential components of the flow have a ratio $E_{\rm pot}/E_{\rm sol} \approx 10^{-4}\mbox{--}10^{-2}$
for most scales; only towards the Nyquist frequency (henceforth $k_{\rm max}$) does the ratio increase to about $0.1$.
It is thus reasonable to assume that compressibility is irrelevant for the results presented here whilst at the same
time the present results can be considered more comparable and relevant to experimental wind tunnel flows than the
perfectly incompressible simulations published so far. The code has been validated in previous turbulence studies and
the reader is referred to Refs.\cite{dhyb03,yhb03,hb04} and the code web-site\cite{PencilCode} for more information.

The simulations are carried out in periodic boxes with resolutions in the range of $32^3 - 512^3$ grid points. The
box size is $L_x = L_y = L_z = 2\pi$, which discretizes the wave numbers in units of 1. The viscosity $\nu$ is
chosen such that the maximum resolved wave number $k_{\rm max}$ is always greater than $1.5/\eta$, where
$\eta=(\nu^3/\varepsilon)^{1/4}$ is the Kolmogorov length scale.

To be consistent with previously published DNS studies, the total kinetic energy $E$ is defined as,
\begin{equation}
E_{\rm tot}=\frac{1}{2}\left \langle \vec{u}^2 \right \rangle = \frac{3}{2}  {u'}^2  = \int_0^{k_{\rm max}}E( k)d
k,\label{eq:e}
\end{equation}
the integral length scale $L$ is defined,
\begin{equation}
L=\frac{\pi}{2   {u'}^2 }\int_0^{k_{\rm max}} k^{-1}E( k)d k,\label{eq:l}
\end{equation}
and the average turbulent energy dissipation rate is defined as
\begin{equation}
 \varepsilon  =2 \nu \int_0^{k_{\rm max}} k^2 E( k)d k.\label{eq:epsilon}
\end{equation}
Angular brackets denote averaging over the box volume. After each run has become statistically stationary
(typically 1-2 eddy turnovers $T \equiv L/ u'$) the average statistics are estimated for the remaining total run
time. Table \ref{tab:average results} summarizes the average statistics for each run. Comparing Runs C and D in
Table \ref{tab:average results} indicates that there is little difference in the average $C_\varepsilon$ for
simulations resolved up to $\eta k_{\rm{max}}=1.5$ from $\eta k_{\rm{max}}=3.$

\section{Results}
\subsection{Numerical results}
\begin{figure}[H]
\begin{center}
\includegraphics[width=0.4\textwidth]{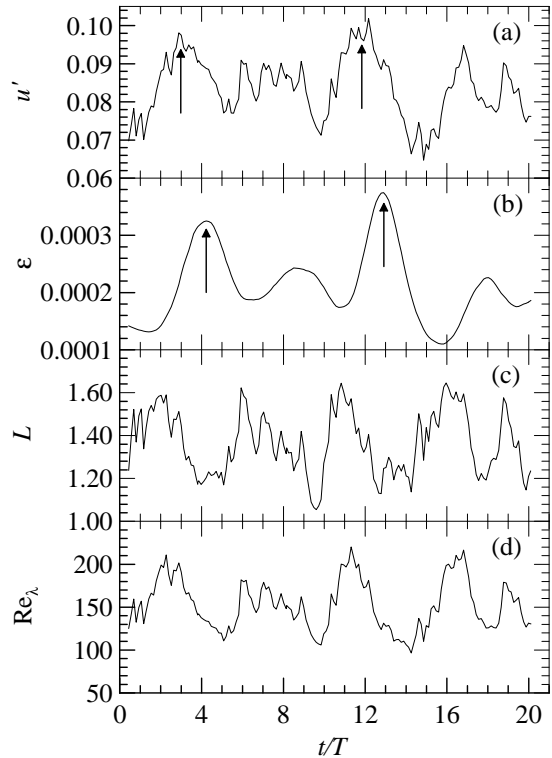}
\caption{Example time series from Run E, $N=256^3$, average $\Rey_\lambda \approx 152$. (a), $u'$; (b),
$\varepsilon$; (c), $L;$ (d), $\Rey_\lambda$. Here, the eddy turnover time $T=L/u'$. The up arrows $\uparrow$
indicate correlated bursts of $u'$ and $\varepsilon.$} \label{fig:ts256}
\end{center}
\end{figure}

In this section results for the higher order finite difference numerical simulations are presented. The simulations
began with $N=32^3$ and each subsequent larger box size began with a velocity field interpolated from the previous box
size. Figures \ref{fig:ts256}(a)-(d) show example time series from Run E $(N=256^3)$ for the fluctuating velocity $u,$
the fluctuating integral length scale $L,$ the fluctuating kinetic energy dissipation rate $\varepsilon$ and the
fluctuating Reynolds number $\Rey_\lambda$ respectively. Initially, the turbulence takes a short amount of time to
reach a statistically stationary state - a consequence of stabilizing the new run from the previously converged run.
The fluctuating quantities shown in Figures \ref{fig:ts256}(a)-(d) are not unlike those encountered in a wind tunnel.
Indeed, Fig.~\ref{fig:ts256}(a) could easily be mistaken for a hot-wire trace of a turbulent flow. This is in stark
contrast to some pseudo-spectral methods that use negative viscosity to maintain a constant energy level. For example,
it is worth comparing Figs.~\ref{fig:ts256}(b)-(d) with those shown in Ref. \cite{ik02} [i.e. their Figs.~(2)-(7)]. The
pseudo spectral results show that the same quantities only fluctuate with a comparatively long period.

\begin{figure}[H]
\begin{center}
\includegraphics[width=0.45\textwidth]{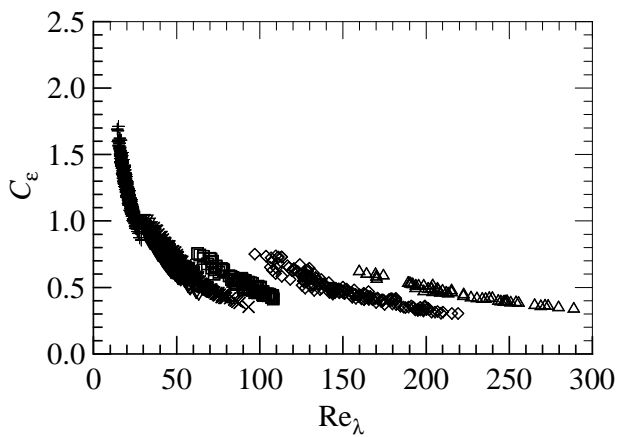}
\caption{Incorrectly estimated $C_\varepsilon$ as a function of $\Rey_\lambda$. $\mathord{+},$ Run A; $\triangledown,$ Run B;
$\times,$ Run C; $\square,$ Run D; $\diamond,$ Run E; $\vartriangle,$ Run F. Ensemble averages can be found in
(Table \ref{tab:average results}).} \label{fig:re_ceps}
\end{center}
\end{figure}

\begin{figure}[t!]
\begin{center}
\includegraphics[width=0.4\textwidth]{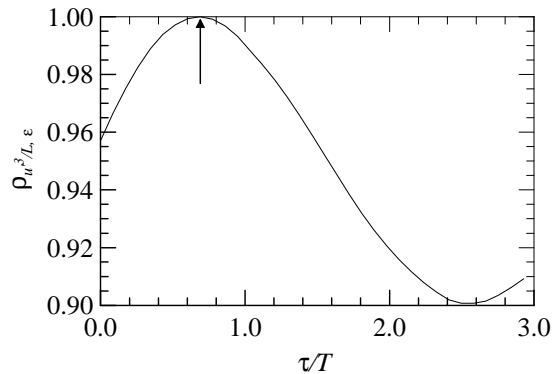}
\caption{An example of the correlation $\rho_{\rm  {u'}^3/L,\varepsilon},$ Eq.~(\ref{eq:rho}), for Run E $N=256^3$.
The up arrow $\uparrow$ indicates the location of $\tau_{\rm max}/T\approx 0.74$.} \label{fig:rho}
\end{center}
\end{figure}

Given that the statistics are fluctuating, although they are statistically stationary, it is tempting to plot the
instantaneous $C_\varepsilon$ as a function of $\Rey_\lambda.$ Figure \ref{fig:re_ceps} shows $C_\varepsilon$
calculated in such a way. The $\Rey_\lambda$ dependent trends are obviously not as expected.
However, it is worth noting the apparent range for $C_\varepsilon$ when $\Rey_\lambda  \gtrsim 50$ is $\approx
0.3-0.7$ which is the range of previously published DNS results. This may explain the scatter in previously
published DNS results if $C_\varepsilon$ is calculated from a subjective choice of $\varepsilon, L $ and $u'$ at a
single time step e.g.\ as in Ref.~\cite{kiyiu03}. The reason for the incorrect $\Rey_\lambda$ dependence for
$C_\varepsilon$ can be gleaned from Figs.~\ref{fig:ts256}(a) and (b). Figure \ref{fig:ts256}(a) shows that an
intense burst in turbulent kinetic energy $u^2$ (an example is noted by the  arrow) can be observed some maximum
time lag $\tau_{\rm max}$ later in the turbulent kinetic energy dissipation rate [Figure \ref{fig:ts256}(b), again
noted by an  arrow].
More about the significance of $\tau_{\rm max}$ will be discussed later in Section~\ref{Experimental}. By noting that there is a
strong correlation between
intense events of $u^2$ and $L$ on the one hand and $\varepsilon$ on the other hand it is possible to estimate
$\tau_{\rm max}$ from the maximum in the correlation between $ {u'}^3/L$ and $\varepsilon$ by
%
\begin{equation}
\label{eq:rho}
\rho_{{u'}^3\!/\!L,\varepsilon} (\tau) = \frac{\overline{[{u'}^3(t)/L(t)]\;[\varepsilon(t+\tau)]}}{\overline{ {u'}^3(t)/L(t)}\; \; \overline{\varepsilon(t+\tau)}}\;,
\end{equation}
Figure \ref{fig:rho} shows an example for Run E. The maximum time lag $\tau_{\rm max}$ corresponding to the maximum in
$\rho_{\rm  {u'}^3\!/L,\varepsilon}$ is indicated by the up arrow $\uparrow$.

With this done for all runs it is possible to shift the time series of $\varepsilon(t)$ for each run by its
respective $\tau_{\rm max}$ and correctly calculate the instantaneous magnitude of $C_\varepsilon.$ Figure
\ref{fig:re_ceps_shift} shows the newly calculated $\Rey_\lambda$ dependence of $C_\varepsilon$ using the correct
time lag $\tau_{\rm max}$ for each of the runs. A number of comments can be made. Firstly, the dimensionless
dissipation rate $C_\varepsilon$ appears to asymptote when $\Rey_\lambda \gtrsim 100.$ The asymptotic magnitude
$C_\varepsilon \approx 0.5$ is in good agreement with the consensus DNS results published so far i.e.\
$C_{\varepsilon} \approx 0.4$ to 0.5. (see Ref. \cite{kiyiu03} and references therein). Having said this and given
the present demonstration that it is incorrect to estimate $C_{\varepsilon}$ from a single time snap shot it would
be interesting to recalculate previously published results based on subjective choices of the quantities involved
for estimating $C_{\varepsilon}$ by using the entire time series. Lastly, the present results verify the use of a
high-order finite difference scheme and also prove that the zeroth law applies to slightly compressible turbulence.

Having estimated $\tau_{\rm max}$ and assuming it approximates the average time $\tau$ for the energy to cascade from
the large energetic scales to the small dissipative scales it is worth comparing the present results with a simple
cascade model such as that discussed by Lumley\cite{l92}. Using a forward cascade model, whereby the spectrum is
divided logarithmically into eddies which have the same width in wave number space as their center wave number, the
total time taken for energy to cross the spectrum, assuming that all energy is passed directly to the next wave number,
\[\tau = \tau_{\rm max}=2\Bigl(\frac L{u'}\Bigr)\Bigl(1-1.29\sqrt{15/[\Rey_{\lambda}^2C_\varepsilon]}\Bigr).\]
Here, we have substituted
$(15/[\Rey_{\lambda}^2C_\varepsilon])^{\frac 1 2}$
for Lumley's large scale Reynolds number dependence
$\ReL_L^{-1/2}.$ In non-dimensional form,
\begin{equation}
\tau\pp = 2\Bigl(1-1.29\sqrt{15/[\Rey_{\lambda}^2C_\varepsilon]}\Bigr). \label{eq:non_dim_time_lag}
\end{equation}
As noted by Lumley, little attention should be paid to the numerical values of the coefficients, though attention
should be paid to the exponent for $\Rey_{\lambda}.$ For small values of $\tau\pp$, e.g.\ $\tau\pp<1$, the large scale
energy is directly affected by viscosity and has little chance of transferring energy in a classical cascade manner,
whilst for large values of $\tau\pp$, e.g.\ $\tau\pp>1$, the large scales have the time for grinding down energy
unaffected by viscosity. The asymptotic assumption of Eq.~(\ref{eq:non_dim_time_lag}) is 2 eddy turnovers. Figure
\ref{fig:non_dim_time_lag}(a) shows the $\Rey_{\lambda}$ dependence of $\tau\pp$ compared with
Eq.~(\ref{eq:non_dim_time_lag}). The present results are much lower than the prediction of
Eq.~(\ref{eq:non_dim_time_lag}) and this is probably indicative of the fact that the energy cascade is not a simple
full transfer of energy between neighboring wave numbers, for low $\Rey_{\lambda}$ at least. It is more likely that,
whilst most of the energy is passed to neighboring wave numbers, a diminishing amount of the energy is passed to all
higher wave numbers. What is noticeable from the present results is that $\tau\pp \approx 1$ will not occur until
$\Rey_{\lambda} \approx 300$ which is an $\Rey_{\lambda}$ at which the $\Rey_{\lambda}$ dependence of $C_\varepsilon$
will become, either numerically or experimentally, unmeasurable. There is no reason not to expect that at high enough
$\Rey_{\lambda}$ full energy transfer may occur between neighboring wave numbers. Using Eq.
(\ref{eq:non_dim_time_lag}), Fig. \ref{fig:non_dim_time_lag} indicates that not until $\Rey_{\lambda}\backsim O(10^3)$
will $\tau\pp \approx 2$.

\subsection{Experimental results revisited}
\label{Experimental}
Results for the present experiment, originally published in Ref.~\cite{pkw02}, are updated here with more data
within the range $170 \lesssim \Rey_{\lambda} \lesssim 1210.$ Detailed experimental conditions can be found in
Refs.~\cite{pkw02,pkj03} and need not be repeated here. The main group of measurements are from a geometry called a
\textsc{Norman} grid which generates a decaying wake flow. The geometry is composed of a perforated plate
superimposed over a bi-plane grid of square rods. The flow cannot be classed as freely decaying as the extent of
the wind tunnel cross section (1.8 $\times$ 2.7 m$^2$) is approximately 7 $\times$ 11 $L^2$. For all the flows
presented in Ref.~\cite{pkw02}, signals of the fluctuating longitudinal velocity $u$ are acquired, for the most
part, on the mean shear profile centerline. For the \textsc{Norman} grid, some data is also obtained slightly off
the center-line at a transverse distance of one mesh height where $dU/dy \approx dU/dy|_{\rm max}/2$.
\begin{figure}[H]
\begin{center}
\includegraphics[width=.8\linewidth]{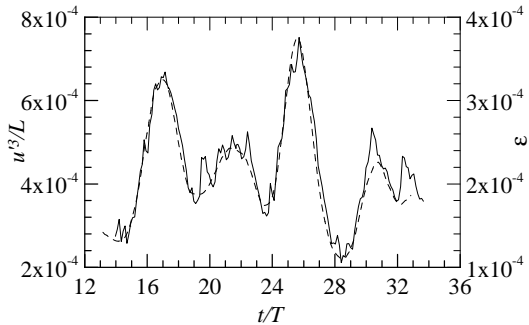}
\caption{Example of the offset time series for Run E ($\tau_{\rm max}\pp\approx 0.74$), $N=256^3,$ average
$\Rey_\lambda \approx 150.$ Note that the peak events are now well correlated. ------, ${u'}^3/L(t/T);$ -- -- --,
$\varepsilon([t-\tau_{\rm max}]/T).$} \label{fig:ts256_shifted}
\end{center}
\end{figure}

All data are acquired using the constant temperature anemometry (CTA) hot-wire technique with a single-wire probe
made of $1.27\mu{\rm m}$\ diameter Wollaston (Pt-10\% Rh) wire. The instantaneous bridge voltage is buck-and-gained and
the amplified signals are low-pass filtered $f_{lp}$\ with the sampling frequency $f_{s}$\ always at least twice
$f_{lp}$. The resulting signal is recorded with 12-bit resolution. Time lags $\tau $ and frequencies $f$ are
converted to streamwise distance $\left(\equiv \tau U\right) $\ and one-dimensional longitudinal wave number $
k_{1}\quad \left( \equiv 2\pi f/U\right) $\ respectively using Taylor's hypothesis. The mean dissipation rate $
\varepsilon  $ is estimated assuming isotropy of the velocity derivatives i.e. $ \varepsilon  \equiv \varepsilon
_{\rm iso} =15\nu  \langle  (
\partial u/\partial x ) ^{2} \rangle.$ We estimate $ \langle  ( \partial u/\partial x )
^{2} \rangle$ from the average value of $E_{\rm 1D}( k_{1})$ [the 1-dimensional energy spectrum of $u$ such that
$ {u}^{2} =\int_{0}^{\infty }E_{\rm 1D}( k_{1})d k_{1}$
and from finite differences $\langle  (\partial u/\partial x ) ^{2} \rangle =  \langle u_{i+1}-u_i
\rangle^2/(Uf_s)^2$].

No corrections for the decrease in wire resolution associated with an increase in $\Rey_{\lambda}$ are made since
all methods known to us rely on an assumed distribution for the three-dimensional energy spectrum. For most of the
data, the worst wire resolution is $\approx2\eta$ where $\eta$ is the dissipative length scale $\equiv
\nu^{3/4}\varepsilon _{\rm iso}^{-1/4}$. The present investigation is limited to one-dimensional measurements and
suitable surrogates for Eq.~(\ref{eq:ceps}). Although caution should be exercised when higher-order moments of a
one-dimensional surrogate are substituted for the three-dimensional equivalent, the use of the mean quantity $
\varepsilon _{\rm iso}$ for $ \varepsilon$ should not be too problematic here. The characteristic length-scale of
the large-scale motions $L$ is $L_{p}$  and is estimated from the wave number $ k_{1,p}$ at which a peak in the
compensated spectrum $ k_{1}E_{\rm 1D}( k_{1})$ occurs i.e. $ L_{p}=1/ k_{1,p}$\cite{b53,l92}. As well the
\textsc{Norman} grid data, the recent cryogenic decaying grid turbulence of White\cite{w02} measured using the
particle image velocimetry (PIV) technique are included.
\begin{figure}[t!]
\begin{center}
\includegraphics[width=0.4\textwidth]{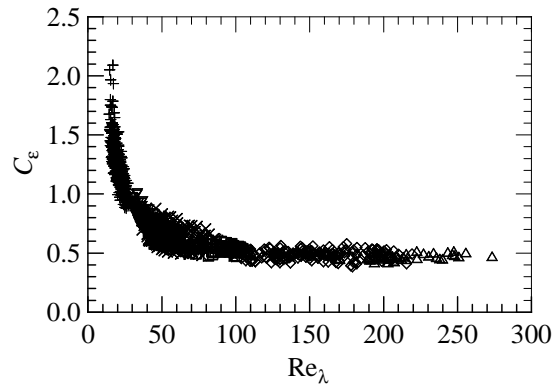}
\caption{Correctly estimated $C_\varepsilon$ as a function of $\Rey_\lambda$. $\mathord +,$ Run A; $\triangledown,$ Run B;
$\times,$ Run C; $\square,$ Run D; $\diamond,$ Run E; $\vartriangle,$ Run F. Ensemble averages can be found in
(Table \ref{tab:average results}).}\label{fig:re_ceps_shift}
\end{center}
\end{figure}
\begin{figure}[H]
\begin{center}
\includegraphics[width=.8\linewidth]{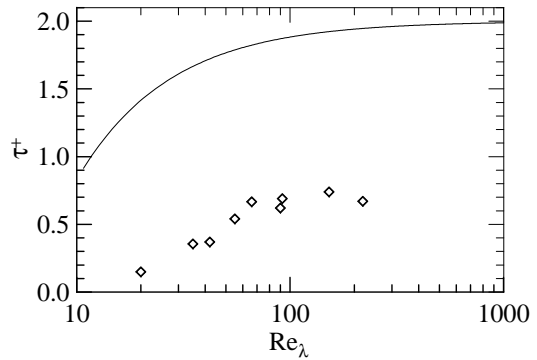}
\caption{$\Rey_\lambda$ dependence of inertial range quantities. $\diamond,$ the non-dimensional time lag $\tau_{\rm
max}\pp = \tau_{\rm max}/T;$ ------, Eq.~(\ref{eq:non_dim_time_lag}).} \label{fig:non_dim_time_lag}
\end{center}
\end{figure}

\begin{figure}[H]
\begin{center}
\includegraphics[width=.8\linewidth]{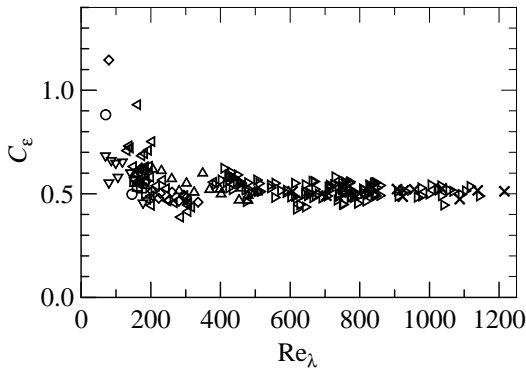}
\caption{ Normalized dissipation rate $C_{\varepsilon}$ for different experimental flows. $\square,$ circular disk,
$154\lesssim \Rey_{\lambda }\lesssim 188; \circ,$ golf ball, $70\lesssim \Rey_{\lambda }\lesssim
146;\triangledown,$ pipe, $70\lesssim \Rey_{\protect\lambda }\lesssim 178 $; $\diamondsuit ,$ normal plate,
$79\lesssim \Rey_{\protect\lambda }\lesssim 335 $; $\vartriangle ,$ \textsc{Norman} grid $N1,152\lesssim
\Rey_{\protect\lambda}\lesssim 506;$ $\times $, \textsc{Norman} grid $N2$ (slight mean shear, $dU/dy \approx
dU/dy|_{max}/2$), $607\lesssim \Rey_{\protect\lambda }\lesssim 1215,  \triangleright,$ \textsc{Norman} grid $N2$
(zero mean shear), $388\lesssim \Rey_{\protect\lambda }\lesssim 1120;$ $ \triangleleft,$ decaying cryogenic grid
turbulence, $127\lesssim \Rey_{\lambda }\lesssim 376$\cite{w02}.} \label{fig:ceps_exp}
\end{center}
\end{figure}

Figure \ref{fig:ceps_exp} shows $C_{\varepsilon}$ for the present data and that of Ref.~\cite{w02}. For all of the
data, a value of $C_{\varepsilon} \approx 0.5$ appears to be the average value. It should be noted that estimates
of $C_{\varepsilon}$ from the cryogenic decaying grid data are based on the transverse equivalents of the
quantities that constitute Eq.~(\ref{eq:ceps}). The majority of the scatter for the cryogenic data is due to the
uncertainty of $L$ which is extremely difficult to estimate from PIV data. Figure \ref{fig:ceps_exp} confirms that
$C_{\varepsilon},$ albeit a one-dimensional surrogate, measured in a number of different flows is independent of
$\Rey_{\lambda}.$ It could be argued that the rate of approach to an asymptotic value depends on the flow e.g.
proximity to initial and boundary conditions. The asymptotic value $C_{\varepsilon}\approx 0.5$ is in excellent
agreement with the present DNS results. These experimental results are encouraging considering that wind-tunnel
turbulence is always relatively young compared to DNS turbulence, e.g. the \textsc{Norman} grid turbulence has only
of the order of 6 eddy turnover times in development by the time it reaches the measurement station.

\section{Final remarks and conclusions}
The present work has revisited the zeroth law of turbulence for both numerical simulations of statistically
stationary isotropic turbulence and experiments. The numerical simulations are slightly compressible isotropic
turbulence and the statistical stationarity is achieved with a random phase forcing applied at low wave numbers.
The main result of the numerical simulations is the demonstration that $C_\varepsilon$ should only be estimated
with ensemble averaged quantities from the entire time series for which the statistics are stationary. If
$C_\varepsilon$ is to be estimated at each time snap shot it is necessary to correctly account for the time lag
that occurs from the large scale energy injection to the fine scale energy dissipation. Even after correctly
correlating the energy injection with the energy dissipation, the instantaneous value of $C_\varepsilon$ can vary
quite considerably (e.g.\ $\pm 30\%$) over the extent of the simulation. Such a variation may account for the
scatter in magnitude of $C_\varepsilon$ in previously published results. Both the present numerical and
experimental results suggest that the asymptotic value for $C_\varepsilon$ is $\approx 0.5.$ In light of this, the
previously held view that the asymptotic value of $C_\varepsilon$ may be dependent on the large scale energy
injection could be suspect. Lastly, the results presented are strictly applicable only to isotropic turbulence that
is stationary in time. However, it would be interesting to estimate $C_\varepsilon$ for simulations of  turbulence
unsteady in space and/or time e.g. anisotropic turbulence or anisotropic homogeneous turbulence with a mean shear
because there is little known for these flows about how the turbulent kinetic energy is dissipated.

\acknowledgments We gratefully acknowledge the Norwegian Research Council for
granting time on the parallel computer in Trondheim (Gridur/Embla) and the NTNU
technical staff for assistance with the experiments.

{\small  CVS ${}$Revision: 1.69 ${}$}
\end{document}